\pdfoutput=1
\documentclass[reprint,aps,prl, twocolumn,floatfix,superscriptaddress]{revtex4-1}
\usepackage{graphicx}
\usepackage{dcolumn}
\usepackage{bm}
\usepackage{amsmath}
\usepackage{amsfonts}
\usepackage{physics}
\usepackage{lineno}
\usepackage[english]{babel}
\usepackage[utf8]{inputenc}
\usepackage{color}
\usepackage{textcomp, gensymb}
\usepackage{upgreek}
\usepackage[
    colorlinks=true,
    citecolor=magenta,
    linkcolor=blue,
    filecolor=magenta,
    urlcolor=magenta,
    hypertexnames=true
]{hyperref}
\renewcommand*{\hbar}{{\mkern-1mu\mathchar'26\mkern-8mu\mathrm{h}}}

\begin{document}

\title{Quantum oscillations in two-dimensional hole gases with competing cyclotron and Zeeman energy}

\author{Davide Costa}
\affiliation{QuTech and Kavli Institute of Nanoscience, Delft University of Technology, Lorentzweg 1, 2628 CJ Delft, Netherlands}
\author{Lucas E. A. Stehouwer}
\affiliation{QuTech and Kavli Institute of Nanoscience, Delft University of Technology, Lorentzweg 1, 2628 CJ Delft, Netherlands}
\author{Davide Degli Esposti}
\affiliation{QuTech and Kavli Institute of Nanoscience, Delft University of Technology, Lorentzweg 1, 2628 CJ Delft, Netherlands}
\author{Giordano Scappucci}
\email{g.scappucci@tudelft.nl}
\affiliation{QuTech and Kavli Institute of Nanoscience, Delft University of Technology, Lorentzweg 1, 2628 CJ Delft, Netherlands}

\date{\today}

\begin{abstract}

Evaluation of critical bandstructure and quantum transport parameters in two-dimensional systems is challenging when competition emerges among different energy scales shaping quantum oscillations in a magnetic field. Here we overcome this challenge in low-disorder strained germanium quantum wells by evaluating self-consistently effective mass, g-factor, and quantum lifetime. As a result, we estimate a quantum mobility of $133(3) \times 10^{3}~\mathrm{cm^2/Vs}$, setting a benchmark for 2D holes in group IV semiconductors. The high quality of the hole gas if further highlighted by observing clean fractional quantum Hall states at low magnetic field and low density.

\end{abstract}

\maketitle

Semiconductor two-dimensional channels are an archetype platform for the development of quantum technology based on spin qubits in quantum dots~\cite{burkard_semiconductor_2023} and superconductor–semiconductor hybrid devices~\cite{prada_andreev_2020}.
The disorder potential in these quantum devices is typically proxied by measurements of charge mobility and percolation-induced critical density of the parent two-dimensional (2D) electron or hole gas.
Other relevant channel properties -- such as out-of-plane effective $g$-factor ($g^*$), in-plane effective mass $m^*$, and quantum lifetime $\tau_\mathrm{q}$~\cite{das_sarma_single-particle_1985} -- are inferred from a careful analysis of the Shubnikov--de Haas oscillations of the magnetoresistivity as a function of temperature $T$ and carrier density $p$~\cite{Coleridge}.
However, the evaluation of these parameters within the validity of the theoretical approximations is challenged by the competition among the energy scales associated with orbital motion, Zeeman spin splitting, spin-orbit interaction, subband splittings, and collisional level broadening. 

This challenge is exemplified in two-dimensional hole gases (2DHG) in strained Ge quantum wells (QWs), recently emerged as a compelling platform for the development of quantum information processing devices~\cite{scappucci2021germanium}. In strained Ge QWs, the valence band ground state has predominantly heavy-hole (HH) symmetry\footnote{cite Winkler, R. Spin-Orbit Coupling Effects in Two-Dimensional Electron and Hole Systems (Springer, 2003).} (pseudo-spin $J_\mathrm{z}= \pm 3/2$), with a remarkably light in-plane effective mass ($\sim0.055m_0$)\cite{lodari2019light,terrazos2021theory} and large out-of-plane $g$-factor of $\sim20$) at the $\Gamma$ point, based on the Luttinger parameters of Ge\cite{Lawaetz1971}.
If the 2DHG carrier density is sufficiently low and close to the $\Gamma$ point, these peculiar valence band states give rise to competing cyclotron energy ($E_{\mathrm{C}} = \frac{\hbar \mathrm{e}}{m^* \mathrm{m_{0}}} B$) and Zeeman energy ($E_{\mathrm{Z}} = g^* \upmu_{\mathrm{B}} B$) even at low magnetic fields $B$.
Previous experimental studies\cite{Lu2017,lu_density-controlled_2017,lodari_lightly_2022,stehouwer2023germanium,myronov2023holes,myronov_electric_2023} have highlighted the unconventional low-density regime where $E_{\mathrm{Z}}$ overcomes the orbital Landau level quantization energy gap $E_{\mathrm{orb}} =E_{\mathrm{C}} - E_{\mathrm{Z}}$. In this regime, SdH oscillations minima for odd (spin-split) filling factors prevail over the even minima, preventing a reliable evaluation of $g^*$ and $\tau_\mathrm{q}$, related to the disorder-induced collisional level broadening $\Gamma = \hbar/2\tau_\mathrm{q}$\cite{dassarma2014mobility}.
In this Letter, we exploit recent advancements in strained Ge QWs with exceptionally low disorder to advance the theoretical framework for evaluating the critical parameters $m^*$, $g^*$, and $\tau_\mathrm{q}$ self-consistently and across the $E_{\mathrm{orb}} \simeq E_{\mathrm{Z}}$ transition, overcoming a long-standing challenge for 2DHGs with HH symmetry.

\begin{figure}
    \centering
	\includegraphics[width=85mm]{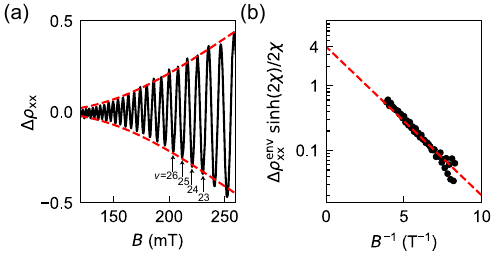}
	\caption{(a) Oscillatory component of the longitudinal resistivity $\Delta \rho_{\mathrm{xx}}$ (solid black curve) at low magnetic fields $B$ after background subtraction. The measurements are performed at a crossover density $p^*$ of $1.28 \times 10^{11} ~\mathrm{cm^{-2}}$, where oscillation minima for even and odd filling factor $\nu$ are observed with equal prominence, implying equal orbital and Zeeman energy gaps. The red dashed curves are the analytical envelope obtained from a single-harmonic approximation, assuming a Lorentzian density of states. (b) Corresponding Dingle plot, where the normalized envelope amplitude $\Delta \rho^{\mathrm{env}}_{\mathrm{xx}} \sinh(2\chi)/2\chi$ is plotted versus $B^{-1}$. The linear fit (red dashed line) yields a quantum lifetime $\tau_{\mathrm{q}} = 3.71(3) ~\mathrm{ps}$ and intercepts the vertical axis at 4, in agreement with the analytical model.}
\label{fig:one}
\end{figure}

\begin{figure*}[t]
    \centering
	\includegraphics[width=160mm]{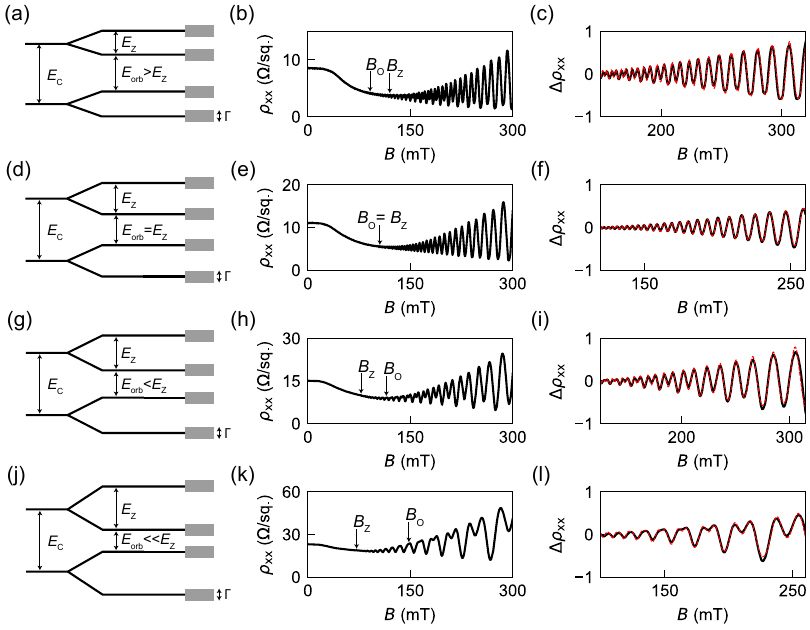}
	\caption{(a),(d),(g),(j) Landau level ladders of the two-dimensional hole gas in a quantizing perpendicular magnetic field, highlighting the interplay between Zeeman and orbital energy for the four measured densities: (a) $p_{\mathrm{2D}} = 1.59 \times 10^{11} ~\mathrm{cm^{-2}}$, $E_\textrm{Z}>E_\textrm{orb}$; (d) $p_{\mathrm{2D}} = 1.28 \times 10^{11} ~\mathrm{cm^{-2}}$, $E_\textrm{Z}=E_\textrm{orb}$; (g) $p_{\mathrm{2D}} = 0.99 \times 10^{11} ~\mathrm{cm^{-2}}$, $E_\textrm{Z}<E_\textrm{orb}$; (j) $p_{\mathrm{2D}} = 0.71 \times 10^{11} ~\mathrm{cm^{-2}}$, $E_\textrm{Z}<<E_\textrm{orb}$. (b),(e),(h),(k) Shubnikov-de Haas oscillations of the longitudinal resistivity $\rho_{\mathrm{xx}}$ at low magnetic field values, corresponding to the energy ladders and density in (a),(d),(g),(j), respectively. $B_\textrm{O}$ and  $B_\textrm{Z}$ mark the magnetic fields for the onset of quantum oscillations corresponding to the orbital and Zeeman gap, respectively. (c),(f),(i),(l) Normalized longitudinal resistivity oscillatory component $\Delta \rho_{\mathrm{xx}}$ (solid black curves) obtained from (a),(d),(g),(j), respectively. Dashed red curves are numerical fits using Eq.\ref{eq1}.}
\label{fig:two}
\end{figure*}

We measure at a temperature $T = 100 ~\mathrm{mK}$ the Ge/SiGe heterostructure field effect transistor studied in Ref~\cite{costa_reducing_2024} with an ultra-high mobility of $4.4(2) \times 10^{6}~\mathrm{cm^2/Vs}$ and a low percolation density of $4.5(1)\times 10^{9} ~\mathrm{cm^{-2}}$.
The low disorder in the channel allows us to measure high-quality low-field SdH oscillations in the longitudinal resistivity $\rho_{\mathrm{xx}}(B)$ down to low density and establish our theoretical framework. We first consider in Fig.~\ref{fig:one}a the oscillatory magnetoresistance $\Delta\rho_{\mathrm{xx}}(B)$\footnote{In our measurements, $\Delta\rho_{\mathrm{xx}}(B)$ is obtained by subtracting a second-degree polynomial background from the normalized longitudinal magnetoresistivity given by
\begin{equation}
\label{eqFOOTNOTE}
    \Delta\rho_{\mathrm{xx}_0}(B) = \frac{\rho_{\mathrm{xx}}(B) - \rho_{\mathrm{xx}}(0)}{\rho_{\mathrm{xx}}(0)} \,.
\end{equation}} measured at a special cross-over density $p^*$
where oscillation minima for even and odd filling factor $\nu$ are observed with equal prominence. In our device, this condition is achieved by fine-tuning the applied gate voltage to set the channel carrier density to $1.28 \times 10^{11} ~\mathrm{cm^{-2}}$. 

Under the assumption of a Lorentzian DOS, and neglecting effects such as spin–orbit coupling, chemical potential oscillations, and non-parabolic band structure, the oscillatory component of the normalized longitudinal resistivity can be expressed as a sum over harmonics $l$, following Ref.~\cite{candido_beating-free_2023}:
\begin{equation}
\begin{split}
\label{eq1}
    \Delta\rho_{\mathrm{xx}}(B) & = 4\sum_{l=1}^{\infty} e^{-l\frac{\pi}{\omega_{\mathrm{c}}\tau_{\mathrm{q}}}} \frac{l\chi}{\sinh \left(l\chi \right)} \times \\
    & \times \cos \left[ 2\pi l \left( \frac{hp}{2eB} -\frac{1}{2} \right) \right] \cos \left( l \frac{\pi}{2} g^* m^* \right) \,,
\end{split}
\end{equation}
where $\omega_{\mathrm{c}} = \mathrm{e}B/m^*\mathrm{m_0}$ is the cyclotron angular frequency and $\chi = 2\uppi^2 \mathrm{k}_{\mathrm{B}} T/\hbar\omega_{\mathrm{c}}$  is the temperature-dependent coefficient. The measurements in Fig.~\ref{fig:one}a imply the condition $E_{\mathrm{orb}} \simeq E_{\mathrm{Z}}$, corresponding to $g^* m^* = 1$, under which the odd harmonics of $\Delta\rho_{\mathrm{xx}}(B)$ in Eq.~\ref{eq1} are suppressed due to the $\cos \left( l \frac{\uppi}{2}\right)$ term.
Since the higher-order harmonics ($l \geq 4$) are exponentially attenuated, we approximate Eq.~\ref{eq1} to the the first even harmonic $l=2$ and obtain for
the envelope of $\Delta\rho_{\mathrm{xx}}$ the expression:
\begin{equation}
\label{eq2}
    \Delta\rho^{\mathrm{env}}_{\mathrm{xx}}(B) \approx 4 e^{-2\frac{\pi }{\omega_{\mathrm{c}}\tau_{\mathrm{q}}}} \frac{2\chi}{\sinh \left(2\chi \right)} \,.
\end{equation}
We use this analytical formula to fit the envelope of the experimental oscillations in Fig.~\ref{fig:one}a assuming, in this first analysis, an effective mass of $0.055m_{\mathrm{0}}$\cite{terrazos2021theory}. This assumption implies  $g^*=1/m* =18.2$ at this special crossover density $p^*$. 
The fit (dashed red lines) shows good agreement with the measured data, supporting the validity of the single-harmonic approximation in this regime.
Figure~\ref{fig:one}b shows the corresponding Dingle plot~\cite{Coleridge} and its linear fit (dashed red line), from which we extract a quantum lifetime $\tau_{\mathrm{q}}$ of $3.71(3) ~\mathrm{ps}$.
Notably, the fit intercepts the $y$ axis at a value of $4$, consistent with the prediction from the analytical expression in Eq.~\ref{eq2}.
This validates our initial assumption of a Lorentzian DOS, confirming our theoretical framework.

We now extend our analysis to the broader range of carrier densities, above or below the crossover density, where the product $g^* m^*$ deviates from unity due to the non-parabolicity of the valence bands and the varying degree of HH-LL mixing.
Figure~\ref{fig:two}a,d,g,j illustrate the energy level diagrams of the 2DHG at four representative densities: $1.59$, $1.28$, $0.99$ and $0.71 \times 10^{11} ~\mathrm{cm^{-2}}$.
The highest density corresponds to the common condition $E_{\mathrm{Z}} < E_{\mathrm{orb}}$, i.e. $g^*m^* < 1$  (Fig.~\ref{fig:two}a).
At the intermediate density $p^*$, the same considered for Fig.~\ref{fig:one}a, the two energy gaps have equal amplitude (Fig.~\ref{fig:two}d) and $g^*m^* = 1$, marking a transition point in the system's energy hierarchy (Fig.~\ref{fig:two}d).
Finally, at the two lowest densities the Zeeman gap exceeds the orbital gap and $g^*m^* > 1$ (Fig.~\ref{fig:two}g-j).
The corresponding SdH oscillations shown in Fig.~\ref{fig:two}b,e,h,k illustrate how the interplay between Zeeman and orbital energy gaps governs the onset and relative strength of quantum oscillations.
To characterize the effects of the competing energy scales, we define $B_{\mathrm{O}}$ and $B_{\mathrm{Z}}$ as the magnetic field values at which the oscillations corresponding to the orbital and Zeeman gap first appear, respectively.
At high density (Fig.~\ref{fig:two}b), where $E_{\mathrm{orb}} > E_{\mathrm{Z}}$, the condition $B_{\mathrm{O}} < B_{\mathrm{Z}}$ is observed: oscillations first emerge from the orbital gap overcoming the disorder broadening $\Gamma$, and even filling factors dominate.
At the crossover density (Fig.~\ref{fig:two}e), $B_{\mathrm{O}} \approx B_{\mathrm{Z}}$, indicating that both gaps contribute equally and oscillations for even and odd filling factors have comparable amplitude.
At lower densities (Fig.\ref{fig:two}h,k), the order is reversed, $B_{\mathrm{O}} > B_{\mathrm{Z}}$: oscillations originate first from the Zeeman gap, and odd filling factors become stronger.
Figure~\ref{fig:two}c,f,i,l shows a first satisfactory theoretical fit of the oscillatory magnetoresistance for the four densities. Similarly to the analysis in Fig.~\ref{fig:one}, we have assumed a constant effective mass $m^*_{i=0} = 0.055\mathrm{m}_{0}$, but differently we use the generalized harmonic expansion of Eq.~\ref{eq1}.

\begin{figure}
  \centering
	\includegraphics[width=85mm]{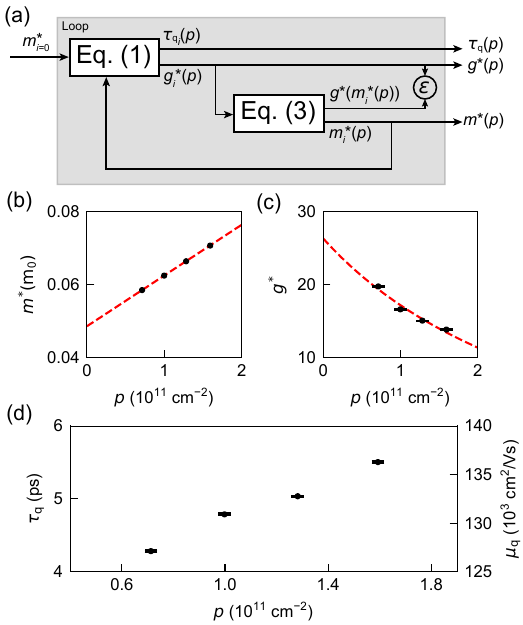}
	\caption{(a) Flowchart of the self-consisting loop for fitting the density-dependent oscillatory component of the magnetoresistance $\Delta \rho_{\mathrm{xx}}$ in Fig.~\ref{fig:two} using the generalized harmonic expansion of Eq.~\ref{eq1}. As an output we obtain the density dependent effective mass $m^*(p)$, effective g-factor $g^*(p)$, and quantum lifetime $\tau_{\mathrm{q}}$. Equation~\ref{eq3} relates $g^*(p)$ and $m^*(p)$, the latter assumed proportional to $p$.   (b) Extracted density-dependent $m^*$ (black dots) with linear fit (red dashed line). (c) Extracted density-dependent $g^*$ (black dots) with theoretical dependence from Eq.~\ref{eq3} (red dashed line). (d) Extracted density-dependent $\tau_{\mathrm{q}}$ (black dots, left axis) and corresponding quantum mobility $\mu_{\mathrm{q}}$ (right axis). Error bars represent fit uncertainties.}
\label{fig:three}
\end{figure}

Building on this first fit, Fig.~\ref{fig:three}a illustrates the self-consistent fitting loop we introduce to extract in a quantitative way the three metrics $g^*$, $m^*$, and $\tau_{\mathrm{q}}$ as a function of density $p$.
Based on previous experimental results~\cite{lodari2019light, drichko2018effective}, we enforce a linear relationship between $m^*$ and $p$ which is then nested into the theoretical relationship that connects $g^*$ and $m^*$ given in Ref.~\cite{drichko2018effective}: 

\begin{equation}
\label{eq3}
\left| g_{\perp} \right| = 2 \left( -3{K} + (\gamma_1 + \gamma_2) - \frac{m_0}{m^*} \right),
\end{equation}
where $\gamma_1 = 13.38$, $\gamma_2 = 4.26$, and $K = 3.41$ are the Luttinger parameters of Ge\cite{Lawaetz1971}.
The loop proceeds as follows: the first iteration, starting with a constant $m^*$, yields preliminary values of $g^*_i(p)$ and $\tau_{\mathrm{q}_i}(p)$ for each density.
Next, these extracted $g^*_i(p)$ values are fitted with the theoretical relation in Eq~\ref{eq3}, from which we obtain a linear dependence of $m^*_i$ on the density.
The updated values $m^*_i(p)$ are then fed back into Eq.\ref{eq1} to refine the fit, producing new sets of values for the parameters $g^*_i(p)$ and $\tau_{\mathrm{q}_i}(p)$.
This procedure is repeated iteratively until the discrepancy $\varepsilon$ between the fitted $g^*_i(p)$ and the mass-dependent $g^*(m^*_i(p))$ is below a threshold of $10^{-2}$.

The outcome of this self-consistent fitting procedure is summarized in Fig.~\ref{fig:three}b,c,d.
The estimated density-dependent effective mass $m^*$  (Fig.~\ref{fig:three}b), with the linear relationship with density imposed a priori, are in line with experiments\cite{lodari2019light} at higher densities and extrapolate to 0.0486(1)$\mathrm{m}_0$ at zero density, in reasonable agreement with theoretical expectations at the $\Gamma$ point\cite{terrazos2021theory}.
The estimated density-dependent $g^*$ (Fig.~\ref{fig:three}b, black points) are in good agreement with the theoretical relationship from Eq.~\ref{eq3} (Fig.~\ref{fig:three}b, red dashed line).
This validation of $g^*$ and $m^*$ with previous reports and theoretical expectations, further reinforces the results obtained for $\tau_{\mathrm{q}}$, shown in Fig.~\ref{fig:three}c along with the quantum mobility $\mu_{\mathrm{q}}=e\tau_{\mathrm{q}}/m^*$.
At the highest density we obtain a remarkable $\tau_{\mathrm{q}}$ of $5.5(1)~\mathrm{ps}$, corresponding to a small energy level broadening $\Gamma=60 \upmu \mathrm{eV}$ and a  $\mu_{\mathrm{q}}$ of $137(1) \times 10^{3}~\mathrm{cm^2/Vs}$.
Differently than transport time and mobility, these metrics are sensitive to scattering through all angles and are indicative of the high quality of the heterostructure\cite{das_sarma_mobility_2014}.
Furthermore, the quantum mobility sets a benchmark for holes in group IV semiconductors, improving previous qualitative estimates in the $25$ to $84 \times 10^{3}~\mathrm{cm^2/Vs}$ range\cite{lodari2021low,myronov_electric_2023,stehouwer2023germanium}.

\begin{figure}[t]
  \centering
	\includegraphics[width=85mm]{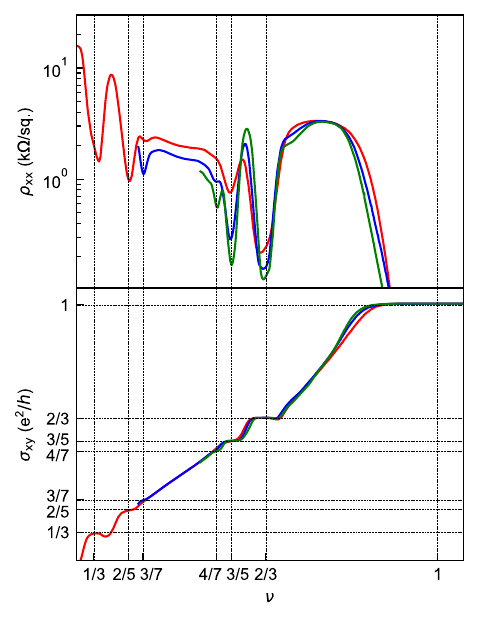}
	\caption{Longitudinal resistivity $\rho_{\mathrm{xx}}$ (top) and normalized transverse conductivity $\sigma_{\mathrm{xy}}$ (bottom) as a function of filling factor $\nu$for $\nu < 1$, measured at three different densities: $0.71 \times 10^{11} ~\mathrm{cm^{-2}}$ (solid red lines), $0.99 \times 10^{11} ~\mathrm{cm^{-2}}$ (solid blue lines) and $1.28 \times 10^{11} ~\mathrm{cm^{-2}}$ (solid green lines). For each curve, the filling factors are obtained via the quantum Hall relationship $\nu = hp/eB$, where $h$ is the Planck constant.}
\label{fig:four}
\end{figure}

The high quantum mobility, indicative of the low-disorder environment in the 2DHG, allows for the observation of an extensive sequence of fractional quantum Hall states at higher magnetic fields, as evidenced by the longitudinal resistivity $\rho_{\mathrm{xx}}$ (top panel) and normalized Hall conductivity $\sigma_{\mathrm{xy}}$ (bottom panel) plotted in Fig.~\ref{fig:four} as functions of the filling factor $\nu$ for three different carrier densities ($1.28$, $0.99$ and $0.71 \times 10^{11} ~\mathrm{cm^{-2}}$).
Pronounced minima in $\rho_{\mathrm{xx}}$ and corresponding quantized plateaus in $\sigma_{\mathrm{xy}}$ are clearly visible at several fractional filling factors of the even denominator series, including $\nu=$1/3, 2/3, 2/5, 3/5, 3/7 and 4/7.
The clear observation of these fractional quantum Hall states at relatively low magnetic fields, with the first fractional state $\nu=$2/3 appearing at fields as low as 5 T for a density of $0.71 \times 10^{11} ~\mathrm{cm^{-2}}$, underscores the remarkable electronic and material quality of the 2DHG.

In summary, we have developed a self-consistent framework to reliably extract the in-plane effective mass, out-of-plane effective $g$-factor, and quantum lifetime in strained Ge quantum wells, overcoming the challenge posed by competing cyclotron and Zeeman energy scales in 2DHGs with heavy-hole symmetry.
Leveraging ultra-low-disorder Ge/SiGe heterostructures, we accessed the unconventional regime where spin splitting dominates over orbital quantization and demonstrated agreement between experiment and theory using a Lorentzian DOS model for disorder broadening.
The high quantum mobility and the observation of clean fractional quantum Hall states highlight the electronic quality of the platform.
Our results establish strained Ge quantum wells as a rich system for exploring spin physics in 2D hole gases, with direct implications for quantum device applications.

\section*{Acknowledgments}
We acknowledge support by the European Union through the IGNITE project with grant agreement No. 101069515 and the QLSI project with grant agreement No. 951852.
This work was supported by the Netherlands Organisation for Scientific Research (NWO/OCW), via the Open Competition Domain Science - M program.
This research was sponsored in part by The Netherlands Ministry of Defence under Awards No. QuBits R23/009. The views, conclusions, and recommendations contained in this document are those of the authors and are not necessarily endorsed nor should they be interpreted as representing the official policies, either expressed or implied, of The Netherlands Ministry of Defence. The Netherlands Ministry of Defence is authorized to reproduce and distribute reprints for Government purposes notwithstanding any copyright notation herein.

\section*{Author Declarations}
 G.S. is founding advisor of Groove Quantum BV and declares equity interests.

\section*{Data availability}
The data sets supporting the findings of this study are openly available in 4TU Research Data at \url{https://doi.org/10.5281/zenodo.17099877}, Ref.~\cite{dataset}.

\end{document}